\documentclass[fleqn,twoside,epsfig]{article}
\usepackage{espcrc2}
\usepackage{amssymb}

\usepackage{graphicx}
\usepackage[figuresright]{rotating}

\newcommand{\be}{\begin{equation}}
\newcommand{\ee}{\end{equation}}

\newcommand{\slh}{\!\!\!\slash}
\newcommand{\psibar}{\overline{\psi}}
\newcommand{\Lqcd}{\Lambda_{\mathrm{QCD}}}

\title{
\vspace{-1.0cm}
\hfill \rm \null \hfill
\hbox{\normalsize ADP-03-127-T563} \\
\vspace{1.0mm}
Scaling Behavior of the Landau Gauge Overlap Quark Propagator 
\thanks{Presented by J. B. Zhang at Lattice 2003}
\thanks{This work is supported by the Australia Research Council}
}
\author{J.\ B.\ Zhang\address[CSSM]{CSSM Lattice Collaboration,\\
Special Research Center for the Subatomic Structure of
Matter (CSSM) and Department of Physics,
University of Adelaide 5005, Australia},
F.\ D.\ R.\ Bonnet\addressmark[CSSM]\address[UR]{Department of Physics, University of Regina, Regina, SK, S4S 0A2, Canada},
P.\ O.\ Bowman\addressmark[CSSM],
D.\ B.\ Leinweber\addressmark[CSSM],
A.\ G.\ Williams\addressmark[CSSM]}
\date{\today}

\begin{document}

\begin{abstract}

The properties of the momentum space quark propagator in Landau gauge
are examined for the overlap quark action in quenched lattice QCD.
Numerical calculations are done on three lattices with different lattice
spacings and similar physical volumes to explore the approach of the quark
propagator towards the continuum limit.  We have calculated the
nonperturbative momentum-dependent wavefunction renormalization function $Z(p^2)$
and the nonperturbative mass function $M(p^2)$ for a variety of bare quark masses
and extrapolate to the chiral limit.
 We find the
behavior of $Z(p^2)$ and $M(p^2)$ are in good agreement
for the two finer lattices in the chiral limit. The quark condensate is 
also calculated.

\end{abstract}

\maketitle
\input epsf

\section{INTRODUCTION}

  There have been several studies of the momentum space quark 
propagator~\cite{jon1,jon2,bowman01,overlgp,over2}
using different gauge fixing and fermion actions. Here we focus on Landau gauge fixing and
the overlap-fermion action and extend previous work~\cite{over2} to three lattices with different lattice
spacings $a$ and very similar physical volumes. This allows us to probe the scaling behavior and the continuum 
limit of the quark propagator in Landau gauge.

\section{QUARK PROPAGATOR ON THE LATTICE}
\label{lattice}
In the continuum the renormalized Euclidean-space
quark propagator must have the form
\begin{eqnarray}
S(\zeta;p)=\frac{1}{i {p \slh} A(\zeta;p^2)+B(\zeta;p^2)}
=\frac{Z(\zeta;p^2)}{i{p\slh}+M(p^2)}\, ,
\label{ren_prop}
\end{eqnarray}
where $Z(\zeta;p^2)$ is the wavefunction renormalization function,  
$M(p^2)$ is the nonperturbative mass function,
and  $\zeta$ is the renormalization point.

On the lattice 
the bare quark propagator  can be written as
\begin{equation}
S^{\rm bare}(p)\equiv
{-i\left(\sum_{\mu}{\cal{C}}_{\mu}(p)\gamma_{\mu}\right)+{\cal{B}}(p)}\, .
\end{equation}
We use periodic boundary conditions in the spatial directions
and anti-periodic in the time direction.
The discrete momentum values for a
lattice of size $N^{3}_{i}\times{N_{t}}$, with $n_i=1,..,N_i$ and $n_t=1,..,N_t$, are
{\small
$$
%\begin{eqnarray}
p_i=\frac{2\pi}{N_{i}a}\left(n_i-\frac{N_i}{2}\right),~~p_t
=\frac{2\pi}{N_{t}a}\left(N_t-\frac{1}{2}-\frac{N_t}{2}\right).
%\label{dismomt}
%\end{eqnarray}
$$
}
We can perform a spinor and color trace to identify
{\small
$$
%\begin{eqnarray}
{\cal{C}}_{\mu}(p)=\frac{i}{12}{\rm tr}[\gamma_\mu{S^{\rm bare}(p)}]
,~~{\cal{B}}(p)=\frac{1}{12}{\rm tr}[S^{\rm bare}(p)] \, .
%\label{curlyCandB}
$$
%\end{eqnarray}
}
The general approach to tree-level
correction\cite{jon2}
utilizes the fact that QCD
is asymptotically free and so it is the difference of bare quantities
from their tree-level form on the lattice that contains the best
estimate of the nonperturbative
information. For the overlap fermion, the tree-level correction is
nothing but to identify appropriate kinematic lattice momentum $q$. 
 We can identify the appropriate kinematic lattice
momentum $q$ directly from the definition of the tree-level quark propagator numerically,
\begin{eqnarray}
q_\mu\equiv C_{\mu}^{(0)}(p)&=&\frac{{\cal{C}}_{\mu}^{(0)}(p)}{({\cal{C}}^{(0)}(p))^{2}+({\cal{B}}^{(0)}(p))^{2}} \, .
\label{latmomt}
\end{eqnarray}
We can also obtain the kinematic lattice momentum $q$ analytically~\cite{overlgp}.

Having identified
the lattice momentum $q$, we can now
define the bare lattice propagator as
\begin{eqnarray}
S^{\rm bare}(p)
&\equiv & \frac{1}{i{q\slh}A(p)+B(p)}
=\frac{Z(p)}{i{q\slh}+M(p)} \nonumber \\
& = & Z_2(\zeta;a) S(\zeta;p) \, ,
\end{eqnarray}
where $S(\zeta;p)$ is 
the lattice version of the renormalized propagator in
Eq.~(\ref{ren_prop}), and $Z_2(\zeta;a)$ is the quark wave-function renormalization constant
chosen so as to ensure $Z(\zeta;\zeta^2)=1$.

The overlap fermion formalism~\cite{neuberger0,neuberger2}
realizes an exact chiral
symmetry on the lattice and is automatically ${\cal O}(a)$ improved.
The massive overlap operator can be written as
as~\cite{edwards2}
\begin{eqnarray}
D(\eta) = \frac{1}{2}\left[1+\eta+(1-\eta)\gamma_5 \epsilon(H)\right] \, ,
\label{D_mu_eqn}
\end{eqnarray}
where $\rho$ is the Wilson mass with negative sign, and the quark mass parameter 
$ \eta \equiv {m_0}/{2\rho}$.
Written according to bare quark mass $m_0$, we have
\begin{equation}
D(m_0)= \frac{1}{2\rho}\left[\rho+\frac{m_0}{2} + (\rho-\frac{m_0}{2})\gamma_5\epsilon(H)\right] \, ,
\end{equation}
and the overlap quark propagator is given by
\begin{equation}
S^{\rm bare}(m_0)\equiv \tilde{D}_c^{-1}(\eta) \, ,
\label{overlap_propagator}
\end{equation}
where
\begin{eqnarray}
\tilde{D}_c^{-1}(\eta)& \equiv& \frac{1}{2\rho} \tilde{D}^{-1}(\eta) \, ,  \\
\tilde{D}^{-1}(\eta) &\equiv& \frac{1}{1-\eta}\left[{D}^{-1}(\eta)-1\right]
\, .
\label{D_mu}
\end{eqnarray}

\section{NUMERICAL RESULTS}
\label{numerical}

Here we work on three lattices with different lattice spacing $a$ and
very similar physical volumes  using a tadpole-improved plaquette plus rectangle
gauge action. For each lattice size, 50 configurations are used.
Lattice parameters are summarized in Table~\ref{simultab}. 

\begin{table}
\caption{Lattice parameters.}
\begin{tabular}{ccccc}
\hline
Action &Volume & $\beta$ &$a$ (fm) &  $u_0$\\
\hline
Improved       & $16^3\times{32}$ & 4.80 & 0.093  & 0.89650 \\
Improved       & $12^3\times{24}$ & 4.60 & 0.125  & 0.88888 \\
Improved       & $8^3\times{16}$  & 4.286& 0.194  & 0.87209 \\
\hline
\end{tabular}
\label{simultab}
\vspace*{-0.5cm}
\end{table}
In the calculations, $\kappa=0.19163$  was used for all three lattices,
which gives $\rho a = (8-1/\kappa)/2 = 1.391$.
We calculate for 10 quark masses on each lattice by using a shifted
Conjugate Gradient solver. The 14th order Zolotarev rational approximation
is used to evaluate the matrix sign function $\epsilon(H_w)$.
The ten bare quark masses we use in our calculation  
are
$m_0=2\rho \eta  =$ $106$, $124$, $142$, $177$, $212$, $266$,
$354$, $442$, $531$, and $620$~MeV respectively.

\begin{figure}[tb]
{\small
%\resizebox*{\columnwidth}{!}{\rotatebox{90}{\includegraphics{./Mext.psq.l12.ps}}}
\resizebox*{\columnwidth}{4.5cm}{\rotatebox{90}{\includegraphics{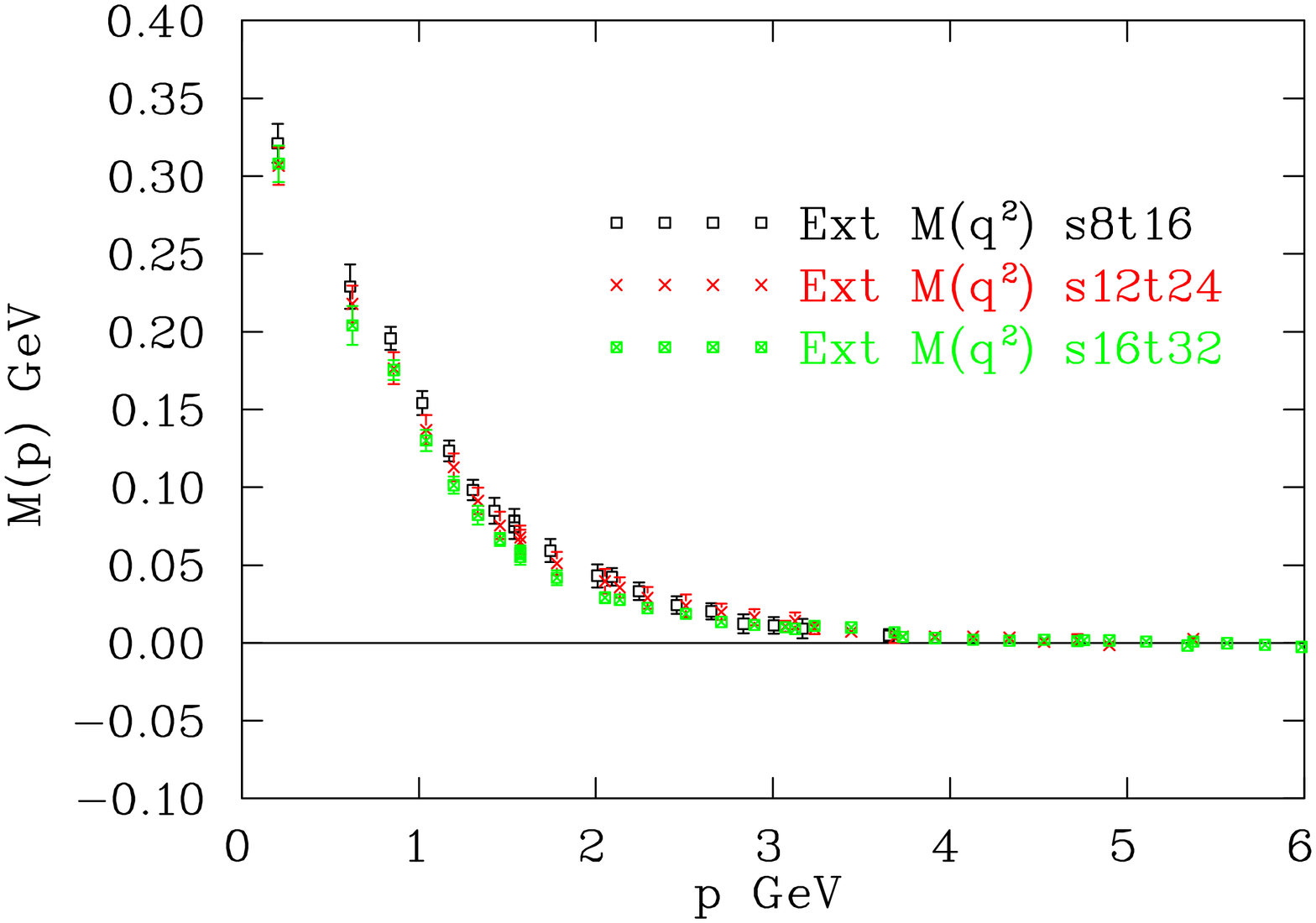}}}
\vspace*{-1.0cm}
\resizebox*{\columnwidth}{4.5cm}{\rotatebox{90}{\includegraphics{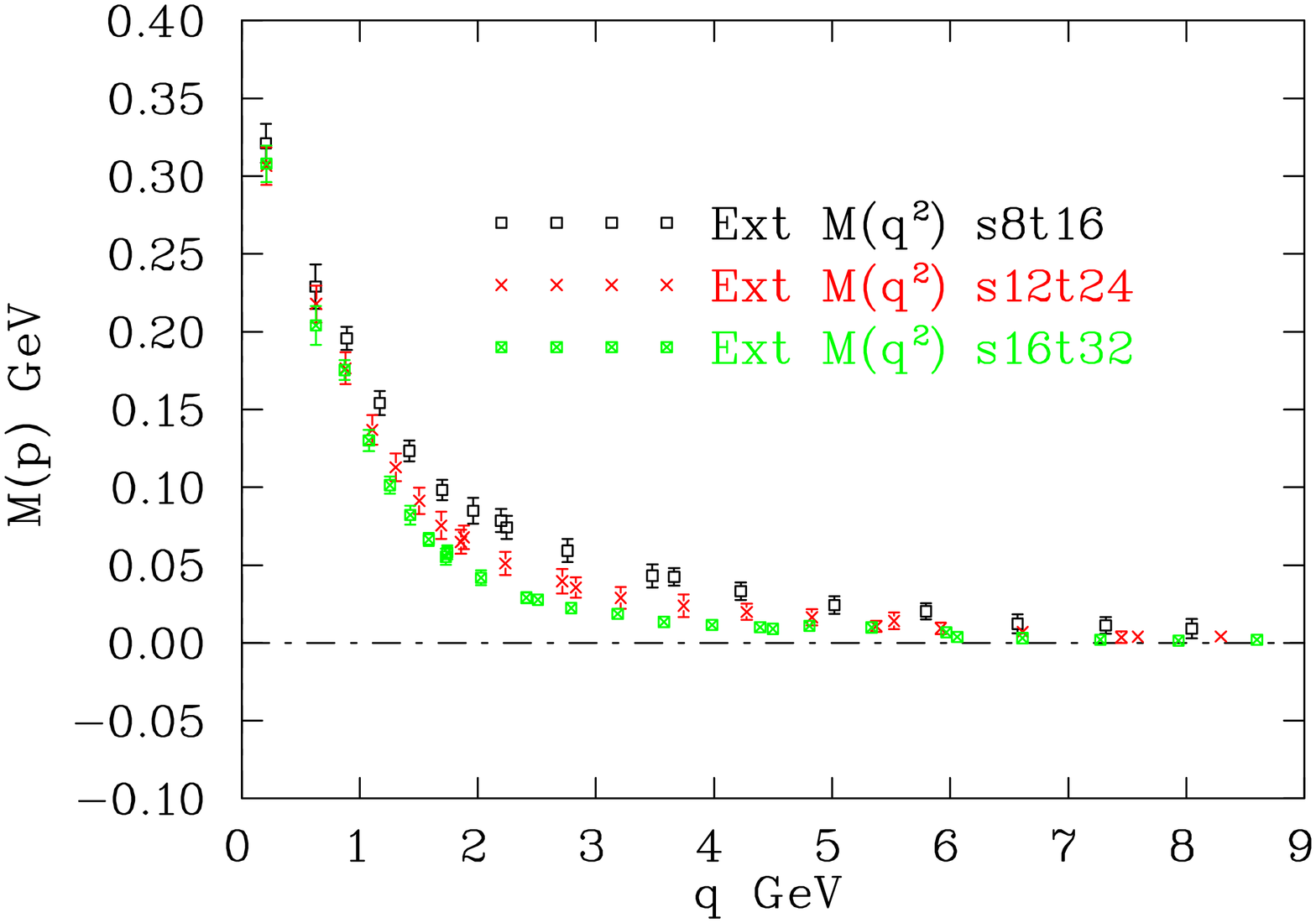}}}
}
\vspace*{-0.5cm}
\caption{\label{ovrmp}{\small Comparison of the mass function $M(p)$ of three lattices in the chiral limit.
The upper graph is plotted against the discrete lattice
momentum $p$ and the lower graph is plotted against the kinematical lattice momentum $q$.}}
\vspace*{-0.6cm}
\end{figure}

\begin{figure}[tb]
\resizebox*{\columnwidth}{4.5cm}{\rotatebox{90}{\includegraphics{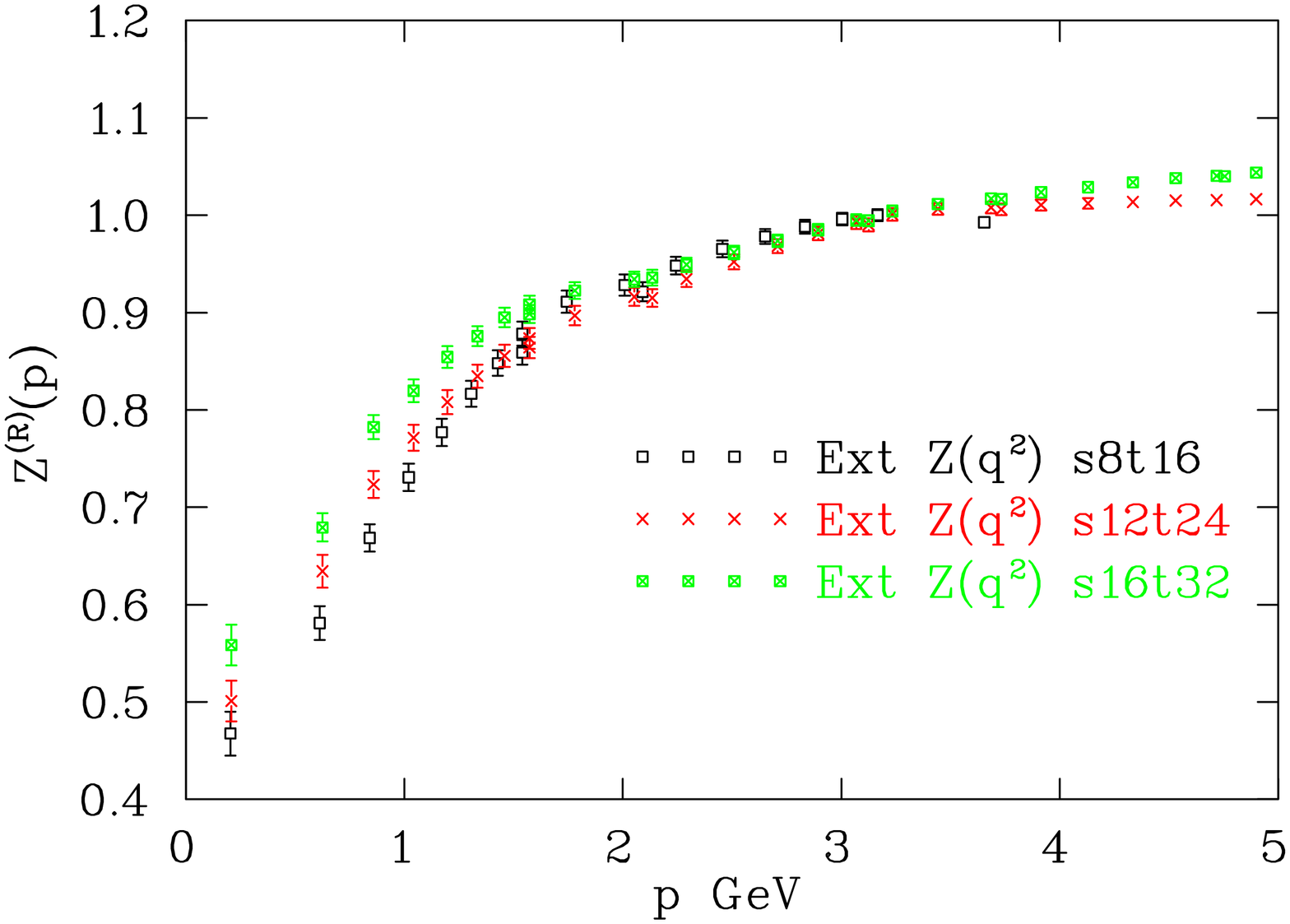}}}
\vspace*{-1.0cm}
\resizebox*{\columnwidth}{4.5cm}{\rotatebox{90}{\includegraphics{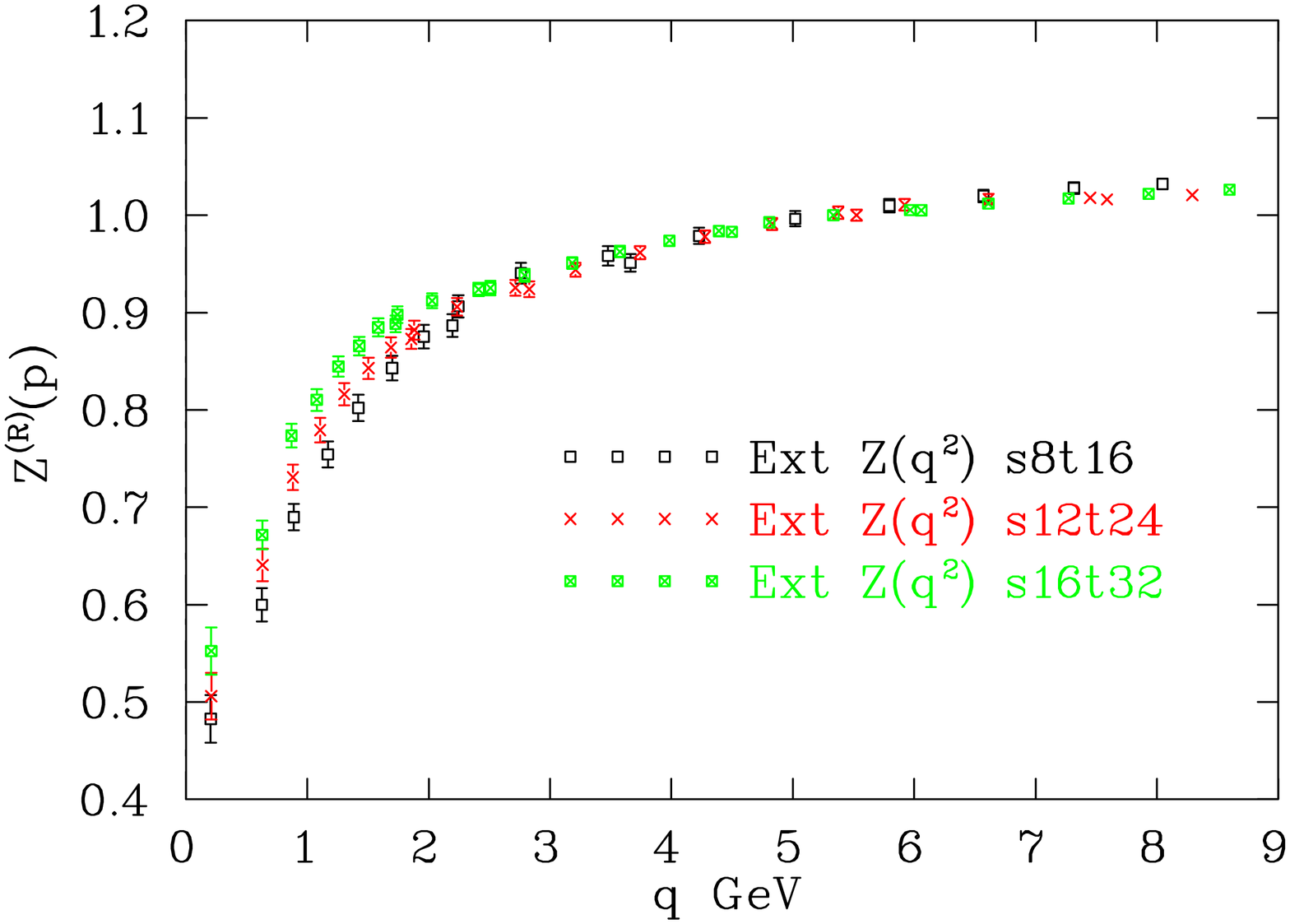}}}
\vspace*{-0.5cm}
\caption{\label{ovrzp}{\small Comparison the renormalization function $Z(p)=Z(\zeta;p)$ for
renormalization point $\zeta$ =5.31 GeV for the $q$-scale and $\zeta$ =3.15 GeV for the $p$-scale 
for the three lattices in the chiral limit. 
The upper graph is plotted against the discrete lattice
momentum $p$ and the lower graph is plotted against the kinematical lattice momentum $q$.}}
\vspace*{-0.8cm}
\end{figure}

 The detailed results will be presented elsewhere. 
Here we focus on the  
comparison of the results on these three lattices.  All data has been cylinder cut~\cite{Leinweber:1998im}
and extrapolated to the chiral limit using a simple linear extrapolation. 
The  mass function $M(p)$ in the chiral limit for the  three lattices is plotted 
in Fig.~\ref{ovrmp} and the renormalization function $Z(p)$ of the three lattices is plotted in Fig.~\ref{ovrzp}.
We can see that when the mass function $M(p)$ is plotted against the discrete lattice momentum $p$
the results of the three lattices are in good agreement,  
while for the renormalization function $Z(p)$, good agreement is reached on the three lattices if it 
is plotted against the kinematical lattice 
momentum $q$. The overall agreement between the two finer lattices is good.

In the chiral and continuum limits, the asymptotic quark mass function has
the form
\begin{eqnarray}
\label{eq:mass_asymp}
M(p^2) \stackrel{p^2\rightarrow \infty}{=} - \frac{4\pi^2 d_M}{3}
   \frac{\langle\psibar \psi\rangle}{[\ln(\mu^2 / \Lqcd^2)]^{d_M}}  \nonumber \\
  \times \frac{[ \ln(p^2 / \Lqcd^2) ]^{d_M-1}}{p^2}
\end{eqnarray}
where the anomalous
dimension of the quark mass is $d_M = 12/(33 - 2N_f)$.
Using the momentum $p$, in the fitting range $ap \subset $(1.3, 2.5), 
on the $16^3 \times 32$ lattice, the resulting value
for the  quark condensate is
\begin{equation}
\langle\psibar \psi\rangle = - (288 \pm 24\mathrm{ MeV})^3.
\end{equation}
This is in excellent agreement with the value $\langle\psibar \psi\rangle = - (268 \pm 27\mathrm{ MeV})^3 $
 extracted from the Asqtad
action using the same method~\cite{bowman02}.

\section{SUMMARY}

 In this report, we have considered tadpole-improved quenched lattice configurations, and  
the overlap fermion operator with the Wilson fermion kernel.
The momentum space quark propagator has been calculated in Landau gauge  
on three lattices with different lattice
spacing $a$ and similar physical volumes to explore the scaling property. 
The continuum limit for $Z(p)$ is most rapidly 
approached when it is plotted against the kinematical lattice momentum $q$, whereas
the quark mass function, $M(p)$, should be plotted against the discrete lattice
momentum $p$. The good agreement between the two finer lattices suggests that we are close to 
the continuum limit.

%\end{references}

\end{document}